\documentclass[a4paper,11pt]{article}
\usepackage[left=3.17cm,right=3.17cm,top=2.54cm,
headheight=0.5cm,headsep=0.54cm,bottom=2.54cm,footskip=0.79cm
]{geometry}

\usepackage{amsmath,amssymb}

\usepackage{tikz}
\usetikzlibrary{arrows}
\usetikzlibrary{decorations.markings}

\usepackage[pdfstartview=FitH]{hyperref}

\begin{document}

\title{A diagrammatic categorification of the fermion algebra}

\author{Bing-Sheng Lin$\,^1$, Zhi-Xi Wang$\,^2$, Ke Wu$\,^2$ and Zi-Feng Yang$\,^2$\\[.3cm]
{\small $^1\,$Department of Mathematics, South China University of Technology,
Guangzhou 510641, China.}\\
{\small $^2\,$School of Mathematical Sciences, Capital Normal University,
Beijing 100048, China.}\\
{\small E-mail: sclbs@scut.edu.cn; wangzhx@cnu.edu.cn; wuke@cnu.edu.cn; yangzf@cnu.edu.cn}
}

\maketitle

\begin{abstract}
In this paper, we study the diagrammatic categorification of the fermion algebra.
We construct a graphical category corresponding to the one-dimensional fermion algebra, and we investigate the properties of this category. The categorical analogues of the Fock states are some kind of 1-morphisms in our category, and the dimension of the vector space of 2-morphisms is exactly the inner product of the corresponding Fock states. All the results in our categorical framework coincide exactly with those in normal quantum mechanics.
\\

\textbf{PACS:} 02.10.Hh, 03.65.Ca, 03.65.Fd

\textbf{Keywords:} categorification, fermion algebra
\end{abstract}

\section{Introduction}\label{sec1}
In general, categorification is a process of replacing set-theoretic theorems by category-theoretic
analogues. It replaces sets by categories, functions by functors, and equations between
functions by natural transformations of functors \cite{Baez}.
The term categorification originated in the work of
Crane and Frenkel on algebraic structures in topological quantum field theories \cite{Crane}.
Categorification can be thought of as the process of enhancing an algebraic
object to a more sophisticated one, while ``decategorification'' is the process of reducing the categorified object back to the simpler original
object. So a useful categorification should possess a richer structure not seen in the underlying object.
In physics, many researchers have introduced category theory into their studies of fundamental physical theories \cite{Baez1}-\cite{Isham}.
The categorification of physical theories may extend the mathematical structures
of existing theories and help us solve the remaining problems in fundamental physics, it can also help us better understand the physical essence.

In recent years, there has been much interest in the studies of the categorification of algebras in mathematical physics \cite{K1}-\cite{Licata}.
The boson algebras (or Heisenberg algebras) and the fermion algebras are the most fundamental algebraic relations in quantum physics.
Recently, Khovanov has constructed a categorification of the Heisenberg algebra based on a graphical category that can act naturally on the category of representations of all symmetric group \cite{Khovanov}, and Licata \emph{et al.} have also done many related works \cite{Cautis, Licata}.
Wang \emph{et al.} also proposed a categorification of the fermions via the categorification of the Heisenberg algebras and the boson-fermion correspondence \cite{Wang}.

In our present work, we use the diagrammatic methods in \cite{Lauda, Khovanov, Licata} to study the categorification of the fermion algebra.
We first construct a graphical category corresponding to the 1D fermion algebra. The states of the fermionic system correspond to some kind of 1-morphisms in this category, and the dimension of the vector space of 2-morphisms is just the inner product of the corresponding states.
Our construction can be easily extended to the case of the higher-dimensional fermion algebras.
Since the fermion algebras can be considered as a special case of the quon algebras when $q=-1$, we also discuss the categorification of the fermion algebras via the categorification of the quon algebras.

This paper is organized as follows. In Section \ref{sec2}, we will briefly review the 1D fermion algebra in normal quantum mechanics. In Section \ref{sec3}, we construct a graphical category that can be used to categorify the 1D fermion algebra and investigate the properties of this category. The categorification of the fermionic Fock states and the categorical inner products are discussed in Section \ref{sec4}.
In Section \ref{sec5}, we construct a 2-representation of our graphical category.
Some discussions are given in Section \ref{sec6}.

\section{The fermion algebra}\label{sec2}
In normal quantum mechanics, the fermionic creation and annihilation operators $\hat{f}^{\dag}$, $\hat{f}$ satisfy the fermion algebraic relations
\begin{eqnarray}\label{1df}
&&\{\hat{f},\hat{f}^{\dag}\}:=\hat{f} \hat{f}^{\dag}+\hat{f}^{\dag} \hat{f}=1,\nonumber\\
&&\{\hat{f},\hat{f}\}=\{\hat{f}^{\dag},\hat{f}^{\dag}\}=0.
\end{eqnarray}
Obviously, we have $\hat{f} \hat{f}=0$ and $\hat{f}^{\dag}\hat{f}^{\dag}=0$.

The Hilbert space is spanned only by two states, $|0\rangle$, $|1\rangle$, where $|0\rangle$ is the vacuum state,
\begin{eqnarray}\label{fs}
&&\hat{f}|0\rangle=0,\qquad \hat{f}^{\dag}|0\rangle =|1\rangle, \nonumber\\
&&\hat{f}|1\rangle =|0\rangle,\qquad \hat{f}^{\dag}|1\rangle =0.
\end{eqnarray}
We have the orthonormal relations
\begin{equation}\label{orth}
\langle n|n'\rangle= \delta_{n,n'}.~~~~(n,n'=0,1)
\end{equation}

The Hamiltonian of the fermionic harmonic oscillator is
\begin{equation}
\hat{H}=\frac{1}{2}(\hat{f}^{\dag}\hat{f}-\hat{f}\hat{f}^{\dag})=\hat{f}^{\dag}\hat{f}-\frac{1}{2},
\end{equation}
here we have set $\hbar=\omega=1$, and we have
\begin{equation}
\hat{H}|0\rangle=-\frac{1}{2}|0\rangle,\qquad \hat{H}|1\rangle=\frac{1}{2}|1\rangle.
\end{equation}

\section{Categorification of the fermion algebra}\label{sec3}
Similar to \cite{Khovanov, Licata}, we may construct an additive $\Bbbk$-linear strict monoidal category $\mathcal{F}$ for a commutative ring $\Bbbk$ as follows. The set of objects in $\mathcal{F}$ is generated by objects $Q_+$ and $Q_-$. An arbitrary object of $\mathcal{F}$ is a finite direct sum of tensor products $Q_\varepsilon := Q_{\varepsilon_1} \otimes
\dots \otimes Q_{\varepsilon_n} $, where $\otimes$ is the ``product'' of the monoidal category, and $\varepsilon = \varepsilon_1
\dots \varepsilon_n$ is a finite sequence of $+$ and $-$ signs.  The
unit object is $\mathbf{1}=Q_\emptyset$.
In this construction, $Q_+$ and $Q_-$ can be regarded as the categorical analogues of the creation and annihilation operators $\hat{f}^{\dag}$, $\hat{f}$ in the previous section.

Similar to \cite{Lauda, Khovanov, Licata}, we may use the string diagrams (that is, planar diagrams) to denote the morphisms in $\mathcal{F}$.
For the properties of the string diagrams, we refer the readers to \cite{Lauda1} for more details.
The space of morphisms $\mathrm{Hom}_{\mathcal{F}}(Q_\varepsilon,
Q_{\varepsilon'})$ is the $\Bbbk$-module generated by string diagrams
modulo local relations.  The diagrams are
oriented compact one-manifolds immersed in the strip $\mathbb{R} \times
[0,1]$, modulo rel boundary isotopies.  The endpoints of the
one-manifold are located at $\{1,\dots,m\} \times \{0\}$ and
$\{1,\dots,k\} \times \{1\}$, where $m$ and $k$ are the lengths of
the sequences $\varepsilon$ and $\varepsilon'$ respectively.  The
orientation of the one-manifold at the endpoints must agree with the
signs in the sequences $\varepsilon$ and $\varepsilon'$, and no triple intersections are allowed.  For example, the diagram
\begin{equation}
\begin{tikzpicture}[>=angle 60,thick,baseline=0pt]
  \draw[<-] (0,.5) to (1,-.5);
  \draw[<-] (0,-.5) to (1,.5);
\end{tikzpicture}
\end{equation}
is one of the morphisms from $Q_{-+}$ to $Q_{+-}$.  A diagram without endpoints gives an endomorphism of $\mathbf{1}$.
The local relations are as follows.

\begin{equation}\label{lcr1}
\begin{tikzpicture}[>=angle 60,thick,baseline=0pt]
  \useasboundingbox (-.2,-1.1) rectangle (7.1,1.1);
  \draw (0,.75) arc (180:360:.5) ;
  \draw (0,1) -- (0,.75) ;
  \draw (1,1) -- (1,.75) [<-];
  \draw (1,-.75) arc (0:180:.5) ;
  \draw (1,-1) -- (1,-.75) ;
  \draw (0,-1) -- (0,-.75) [<-];
  \draw (1.5,0) node {$=$};
  \draw (2,-1) -- (2,1) [<-];
  \draw (3,-1) -- (3,1) [->];
  \draw (3.5,0) node {,};
  \draw (5,-1) -- (5,1) [->];
  \draw (6,-1) -- (6,1) [->];
  \draw (6.5,0) node {$=$};
  \draw (7,0) node {$0$};
\end{tikzpicture}
\end{equation}
\ \\
\begin{equation}\label{lcr2}
\begin{tikzpicture}[>=angle 60,thick,baseline=0pt,
decoration={markings,mark=at position 0 with {\arrowreversed[]{angle 60}}}]
  \draw [postaction={decorate}] (.5,0) circle (.5);
  \draw (1.5,0) node {$+$};
  \draw[xscale=-1,shift={(-5,0)}] [postaction={decorate}] (2.5,0) circle (.5);
  \draw (3.5,0) node {$=$};
  \draw (4,0) node {$\mathrm{id}$};
  \useasboundingbox (-.1,-.6) rectangle (4.1,.6);
\end{tikzpicture}
\end{equation}
\ \\
The second relation in \eqref{lcr1} means that $Q_{++}\cong \mathbf{0}$, where $\mathbf{0}$ is zero object in the additive category $\mathcal{F}$. In a linear category, $\mathrm{id}_{A}=0$ for an object $A$ implies that $A$ is isomorphic to the zero object. We also have $Q_{--}\cong\mathbf{0}$. These just correspond to $(\hat{f}^{\dag})^2=0$ and $(\hat{f})^2=0$ in the previous section.
Note that, from the first relation in \eqref{lcr1} we may obtain the relation
\begin{equation}
\begin{tikzpicture}[>=angle 60,thick,baseline=0pt]
  \draw (0,.75) arc (180:360:.5) ;
  \draw (0,1) -- (0,.75) [<-];
  \draw (1,1) -- (1,.75);
  \draw (1,-.75) arc (0:180:.5) ;
  \draw (1,-1) -- (1,-.75) [<-];
  \draw (0,-1) -- (0,-.75);
  \draw (1.5,0) node {$=$};
  \draw (2,-1) -- (2,1) [->];
  \draw (3,-1) -- (3,1) [<-];
  \useasboundingbox (-.2,-1.1) rectangle (3.2,1.1);
\end{tikzpicture}
\end{equation}
and vice versa.

We find that the local relations of the fermion algebra are much simpler than those of the boson algebra \cite{Khovanov,Licata}.

From the local relations above, we will see that in the category $\mathcal{F}$ we have the following isomorphic relation
\begin{equation}\label{fa}
  Q_{+-} \oplus Q_{-+} \cong \mathbf{1}.
\end{equation}
Consider the following morphisms of $\mathcal{F}$,
\begin{equation}
\begin{tikzpicture}[>=angle 60,thick,baseline=0pt]
  \draw (0,3) node {$\mathbf{1}$};
  \draw (0,-3) node {$\mathbf{1}$};
  \draw (-3,0) node {$Q_{+-}$};
  \draw (3,0) node {$Q_{-+}$};
  \draw[->] (-.5,-2.5) to (-2.5,-.5);
  \draw (-1.2,-1.2) node {$\rho_1$};
  \draw[->] (.5,-2.5) to (2.5,-.5);
  \draw (1.2,-1.2) node {$\rho_2$};
  \draw[->] (-2.5,.5) to (-.5,2.5);
  \draw (-1.2,1.2) node {$\iota_1$};
  \draw[->] (2.5,.5) to (.5,2.5);
  \draw (1.2,1.2) node {$\iota_2$};
  \draw[<-] (1.6,1.7) arc(180:0:.5);
  \draw[->] (1.6,-1.7) arc(180:360:.5);
  \draw[->] (-2.6,1.7) arc(180:0:.5);
  \draw[<-] (-2.6,-1.7) arc(180:360:.5);
\end{tikzpicture}
\end{equation}
from the defining local relations \eqref{lcr1} and \eqref{lcr2} of $\mathcal{F}$, we have
\begin{equation}
  \rho_2 \iota_1 = 0,\quad \rho_1 \iota_2 = 0,\quad \rho_1 \iota_1 = \mathrm{id},\quad \rho_2 \iota_2 = \mathrm{id},\quad \iota_1 \rho_1 + \iota_2 \rho_2 = \mathrm{id},
\end{equation}
this proves the isomorphism \eqref{fa}.

So in the Grothendieck group $K_0(\mathcal{F})$, we have
\begin{equation}
  [Q_+][Q_-] + [Q_-][Q_+] = 1,
\end{equation}
which is the fermion algebraic relation \eqref{1df}.

From the second relation in \eqref{lcr1}, we find that \emph{all the string diagrams with crossings are equal to zero}.
So the string diagrams in the category $\mathcal{F}$ are much simpler than those in the graphical categories corresponding to the boson algebra \cite{Khovanov, Licata}.
We also have the following relation
\begin{equation}
\begin{tikzpicture}[>=angle 60,thick,baseline=0pt]
  \draw [postaction={decoration={markings,mark=at position 0 with {\arrowreversed[]{angle 60}}}, decorate}] (-.5,0) circle (.5);
  \draw[->] (0.5,-1) to (0.5,1);
  \draw (1.25,0) node {$=$};
  \draw[->] (2,-1) to (2,1);
  \draw (2.5,0) node {,};
  \draw[yscale=-1] [postaction={decoration={markings,mark=at position 0 with {\arrowreversed[]{angle 60}}}, decorate}] (4.5,0) circle (.5);
  \draw[->] (5.5,-1) to (5.5,1);
  \draw (6.5,0) node {$=~~~0$};
  \useasboundingbox (-1.1,-1.1) rectangle (6.6,1.1);
\end{tikzpicture}
\end{equation}
It is easy to see that if there are some other lines in a diagram, then the bubbles can be omitted, or the whole diagram might be $0$.

Obviously, any non-zero object can be expressed as a finite direct sum of $Q_{-+\,\cdots\,-+}$, $Q_{+-+\,\cdots\,-+}$, $Q_{+-\,\cdots\,+-}$ and $Q_{-+-\,\cdots\,+-}$,
so a basis of the $\Bbbk$-vector space $\mathrm{Hom}_{\mathcal{F}}(Q_{\varepsilon},Q_{\varepsilon'})$ for $\varepsilon, \varepsilon'\neq \emptyset$ consists of one of the following string diagrams

\begin{equation*}
\begin{tikzpicture}[>=angle 60,thick,baseline=0pt,scale=0.6]
    \draw (1,1) arc(180:360:.5);
    \draw (1,1) to (1,2);
    \draw[->] (2,1) to (2,2);
    \draw (3,1) arc(180:360:.5);
    \draw (3,1) to (3,2);
    \draw[->] (4,1) to (4,2);
    \draw (3,-1) arc(180:0:.5);
    \draw[->] (3,-1) to (3,-2);
    \draw (4,-1) to (4,-2);
    \draw[loosely dotted,line width=1pt] (4.5,1.5) to (5.5,1.5);
    \draw[loosely dotted,line width=1pt] (4.5,-1.5) to (5.5,-1.5);
    \draw (6,1) arc(180:360:.5);
    \draw (6,1) to (6,2);
    \draw[->] (7,1) to (7,2);
    \draw (6,-1) arc(180:0:.5);
    \draw[->] (6,-1) to (6,-2);
    \draw (7,-1) to (7,-2);
    \draw (8.5,0) node {,};
    \draw (11,1) arc(180:360:.5);
    \draw (11,1) to (11,2);
    \draw[->] (12,1) to (12,2);
    \draw[->] (12,-2) .. controls (12,-.5) .. (11,0) .. controls (10,.5) .. (10,2);
    \draw (13,1) arc(180:360:.5);
    \draw (13,1) to (13,2);
    \draw[->] (14,1) to (14,2);
    \draw (13,-1) arc(180:0:.5);
    \draw[->] (13,-1) to (13,-2);
    \draw (14,-1) to (14,-2);
    \draw[loosely dotted,line width=1pt] (14.5,1.5) to (15.5,1.5);
    \draw[loosely dotted,line width=1pt] (14.5,-1.5) to (15.5,-1.5);
    \draw (16,1) arc(180:360:.5);
    \draw (16,1) to (16,2);
    \draw[->] (17,1) to (17,2);
    \draw (16,-1) arc(180:0:.5);
    \draw[->] (16,-1) to (16,-2);
    \draw (17,-1) to (17,-2);
    \useasboundingbox (.9,-2.1) rectangle (17.2,2.1);
  \end{tikzpicture}
\end{equation*}

\ \\

\begin{equation}
\begin{tikzpicture}[>=angle 60,thick,baseline=0pt,scale=0.6]
    \draw (1,1) arc(180:360:.5);
    \draw[->] (1,1) to (1,2);
    \draw (2,1) to (2,2);
    \draw (3,1) arc(180:360:.5);
    \draw[->] (3,1) to (3,2);
    \draw (4,1) to (4,2);
    \draw (3,-1) arc(180:0:.5);
    \draw (3,-1) to (3,-2);
    \draw[->] (4,-1) to (4,-2);
    \draw[loosely dotted,line width=1pt] (4.5,1.5) to (5.5,1.5);
    \draw[loosely dotted,line width=1pt] (4.5,-1.5) to (5.5,-1.5);
    \draw (6,1) arc(180:360:.5);
    \draw[->] (6,1) to (6,2);
    \draw (7,1) to (7,2);
    \draw (6,-1) arc(180:0:.5);
    \draw (6,-1) to (6,-2);
    \draw[->] (7,-1) to (7,-2);
    \draw (8.5,0) node {,};
    \draw (11,1) arc(180:360:.5);
    \draw[->] (11,1) to (11,2);
    \draw (12,1) to (12,2);
    \draw[<-] (12,-2) .. controls (12,-.5) .. (11,0) .. controls (10,.5) .. (10,2);
    \draw (13,1) arc(180:360:.5);
    \draw[->] (13,1) to (13,2);
    \draw (14,1) to (14,2);
    \draw (13,-1) arc(180:0:.5);
    \draw (13,-1) to (13,-2);
    \draw[->] (14,-1) to (14,-2);
    \draw[loosely dotted,line width=1pt] (14.5,1.5) to (15.5,1.5);
    \draw[loosely dotted,line width=1pt] (14.5,-1.5) to (15.5,-1.5);
    \draw (16,1) arc(180:360:.5);
    \draw[->] (16,1) to (16,2);
    \draw (17,1) to (17,2);
    \draw (16,-1) arc(180:0:.5);
    \draw (16,-1) to (16,-2);
    \draw[->] (17,-1) to (17,-2);
    \useasboundingbox (.8,-2.1) rectangle (17.2,2.1);
  \end{tikzpicture}
\end{equation}
We also have the diagrams
\begin{equation}
\begin{tikzpicture}[>=angle 60,thick,baseline=0pt,
decoration={markings,mark=at position 0 with {\arrowreversed[]{angle 60}}}]
  \draw [postaction={decorate}] (-2,0) circle (.5);
  \draw (-.5,0) node {,};
  \draw[xscale=-1] [postaction={decorate}] (-2,0) circle (.5);
  \useasboundingbox (-2.6,-.6) rectangle (2.6,.6);
\end{tikzpicture}
\end{equation}
for $\mathrm{Hom}_{\mathcal{F}}(Q_{\emptyset},Q_{\emptyset})$ or $\mathrm{Hom}_{\mathcal{F}}(\mathbf{1},\mathbf{1})$.

In fact, using the string diagrams, it is easy to see that we have the following isomorphic relations $Q_{-+\,\cdots\,-+}\cong Q_{-+}$\,, $Q_{+-\,\cdots\,+-}\cong Q_{+-}$.
We also have $Q_{-+\,\cdots\,-+-}\cong Q_{-}$\,, $Q_{+-\,\cdots\,+-+}\cong Q_{+}$.
For example, for $Q_{+-+-}\cong Q_{+-}$, we have the following morphisms,
\begin{equation}
\begin{tikzpicture}[>=angle 60,thick,baseline=0pt,scale=0.7]
    \draw[<-] (0,2) .. controls (0,1) and (1,1) .. (1,2);
    \draw[<-] (2,2) .. controls (2,1) and (3,1) .. (3,2);
    \draw[->] (2,0) .. controls (2,1) and (3,1) .. (3,0);
    \draw[->] (0,-2) .. controls (0,-1) and (1,-1) .. (1,-2);
    \draw[->] (2,-2) .. controls (2,-1) and (3,-1) .. (3,-2);
    \draw[<-] (2,0) .. controls (2,-1) and (3,-1) .. (3,0);
    \draw (4,0) node {$=$};
    \draw[<-] (5,2) to (5,-2);
    \draw[->] (6,2) to (6,-2);
    \draw[<-] (7,2) to (7,-2);
    \draw[->] (8,2) to (8,-2);
    \draw (9,0) node {,};
    \draw[<-] (11,0) .. controls (11,-1) and (12,-1) .. (12,0);
    \draw[<-] (13,0) .. controls (13,-1) and (14,-1) .. (14,0);
    \draw[->] (13,-2) .. controls (13,-1) and (14,-1) .. (14,-2);
    \draw[->] (11,0) .. controls (11,1) and (12,1) .. (12,0);
    \draw[->] (13,0) .. controls (13,1) and (14,1) .. (14,0);
    \draw[<-] (13,2) .. controls (13,1) and (14,1) .. (14,2);
    \draw (15,0) node {$=$};
    \draw[<-] (16,2) to (16,-2);
    \draw[->] (17,2) to (17,-2);
    \useasboundingbox (-.2,-2.1) rectangle (17.2,2.1);
  \end{tikzpicture}
\end{equation}

Similar to \cite{Licata}, we may define the 2-category $\mathfrak{F}$ as follows.
There are only two objects ``$0$'' and ``$1$'' in the 2-category $\mathfrak{F}$.
For $n,m \in \mathrm{Ob}\mathfrak{F}$, $\mathrm{Hom}_{\mathfrak{F}}(n,m)$ is the full
subcategory of $\mathcal{F}$ containing the objects $Q_\varepsilon$,
$\varepsilon = \varepsilon_1 \dots \varepsilon_l$, for which
\[
  m-n = \# \{i\ |\ \varepsilon_i = +\} - \# \{i\ |\ \varepsilon_i =
  -\}.
\]
In this case, the regions of the string diagrams are labelled by $0$ or $1$, any string diagram which has a region labelled by some number less than $0$ or greater than $1$ will be set to zero. For example, the following string diagrams are equal to zero,
\begin{equation}
\begin{tikzpicture}[>=angle 60,thick,baseline=0pt,
decoration={markings,mark=at position 0 with {\arrowreversed[]{angle 60}}}]
  \draw [postaction={decorate}] (-2.5,0) circle (.5);
  \draw (-2.5,0) node {$-1$};
  \draw (-1.5,0) node {$0$};
  \draw (-.5,0) node {,};
  \draw[xscale=-1] [postaction={decorate}] (-2.5,0) circle (.5);
  \draw (2.5,0) node {$2$};
  \draw (3.5,0) node {$1$};
  \useasboundingbox (-3.1,-.6) rectangle (3.6,.6);
\end{tikzpicture}
\end{equation}
So in the 2-category $\mathfrak{F}$, the local relations \eqref{lcr1}, \eqref{lcr2} will become
\begin{equation}\label{2lr1}
\begin{tikzpicture}[>=angle 60,thick,baseline=0pt]
  \draw (0,.75) arc (180:360:.5) ;
  \draw (0,1) -- (0,.75) ;
  \draw (1,1) -- (1,.75) [<-];
  \draw (1,-.75) arc (0:180:.5) ;
  \draw (1,-1) -- (1,-.75) ;
  \draw (0,-1) -- (0,-.75) [<-];
  \draw (1.5,0) node {$0$};
  \draw (2.25,0) node {$=$};
  \draw (3,-1) -- (3,1) [<-];
  \draw (4,-1) -- (4,1) [->];
  \draw (4.5,0) node {$0$};
  \draw (5.5,0) node {,};
  \draw[yscale=-1] [postaction={decoration={markings,mark=at position 0 with {\arrowreversed[]{angle 60}}}, decorate}] (7.5,0) circle (.5);
  \draw (8.5,0) node {$0$};
  \draw (9.25,0) node {$=$};
  \draw (10,0) node {$\mathrm{id}_{\mathrm{id}_{0}}$};
  \useasboundingbox (-.2,-1.1) rectangle (10.2,1.1);
\end{tikzpicture}
\end{equation}
\begin{equation}\label{2lr2}
\begin{tikzpicture}[>=angle 60,thick,baseline=0pt]
  \draw (0,.75) arc (180:360:.5) ;
  \draw (0,1) -- (0,.75) [<-];
  \draw (1,1) -- (1,.75);
  \draw (1,-.75) arc (0:180:.5) ;
  \draw (1,-1) -- (1,-.75) [<-];
  \draw (0,-1) -- (0,-.75);
  \draw (1.5,0) node {$1$};
  \draw (2.25,0) node {$=$};
  \draw (3,-1) -- (3,1) [->];
  \draw (4,-1) -- (4,1) [<-];
  \draw (4.5,0) node {$1$};
  \draw (5.5,0) node {,};
  \draw [postaction={decoration={markings,mark=at position 0 with {\arrowreversed[]{angle 60}}}, decorate}] (7.5,0) circle (.5);
  \draw (8.5,0) node {$1$};
  \draw (9.25,0) node {$=$};
  \draw (10,0) node {$\mathrm{id}_{\mathrm{id}_{1}}$};
  \useasboundingbox (-.2,-1.1) rectangle (10.2,1.1);
\end{tikzpicture}
\end{equation}
and the diagram of the second relation in \eqref{lcr1} with any labels in the regions is naturally equal to zero.

Furthermore, from the relations \eqref{2lr1}, \eqref{2lr2}, it is easy to see that we have the following isomorphic relations
\begin{equation}
Q_{-+}\cong \mathrm{id}_{0}\,,\qquad Q_{+-}\cong \mathrm{id}_{1}\,,
\end{equation}
these are just the categorical analogues to the relations
\begin{equation}
\hat{f}\hat{f}^{\dag}|0\rangle=|0\rangle, \qquad
\hat{f}^{\dag}\hat{f}|1\rangle=|1\rangle.
\end{equation}

\section{Categorification of the Fock states and the inner product}\label{sec4}

In the 2-category $\mathfrak{F}$, the 1-morphisms $Q_+$, $Q_-$ correspond to the operators $\hat{f}^{\dag}$, $\hat{f}$ respectively. We now introduce the notation $\hat{A}_\varepsilon := \hat{f}_{\varepsilon_1}\dots \hat{f}_{\varepsilon_n}$, where $\varepsilon = \varepsilon_1
\dots \varepsilon_n$ is a finite sequence of $+$ and $-$ signs, and $\hat{f}_{+}:=\hat{f}^{\dag}$, $\hat{f}_{-}:=\hat{f}$.
We know in Fock space, for any state $|\psi\rangle$, we have $|\psi\rangle=f(\hat{f},\hat{f}^{\dagger})|0\rangle$, where $f(\hat{f},\hat{f}^{\dagger})$ is some function of the operators $\hat{f}$, $\hat{f}^{\dagger}$. So any state $|\psi\rangle$ can be uniquely determined by the corresponding operator $f(\hat{f},\hat{f}^{\dagger})$.
Similar to \cite{lw}, a non-zero Fock state $|\hat{A}_\varepsilon\rangle:=\hat{A}_\varepsilon|0\rangle=\lambda|n\rangle$ for $\lambda\in\mathbb{R}$ corresponds to some 1-morphism $Q_\varepsilon : 0 \to n$ for $n=0,1$. The inner product $\langle\hat{A}_{\varepsilon'}|\hat{A}_{\varepsilon}\rangle$ corresponds to the dimension of the vector space $\mathrm{Hom}_{\mathfrak{F}}(Q_\varepsilon,Q_{\varepsilon'})$ for the corresponding 1-morphisms $Q_\varepsilon$ and $Q_{\varepsilon'}$ \cite{Lauda, KMS},
\begin{equation}
\langle\hat{A}_{\varepsilon'}|\hat{A}_{\varepsilon}\rangle
=\langle Q_{\varepsilon'}, Q_{\varepsilon}\rangle
:=\mathrm{dim(Hom}_{\mathfrak{F}}(Q_\varepsilon, Q_{\varepsilon'})).
\end{equation}

So the states $|0\rangle$, $|1\rangle$ correspond to the 1-morphisms $\psi_0:=Q_{-+}:0\to 0$, $\psi_1:=Q_{+}:0\to 1$, respectively. Since we have $\psi_0\cong Q_{-+\,\cdots\,-+}\cong \mathrm{id}_{0}$, $\psi_1\cong Q_{+-+\,\cdots\,-+}$,
\emph{these categorical states are uniquely determined up to isomorphism}.
Obviously, we may also regard the objects ``$0$'', ``$1$'' in our category as the states $|0\rangle$, $|1\rangle$, but it is more convenient to consider the 1-morphisms as the states, and this is also the spirit in the framework of categorical quantum mechanics developed by Abramsky and Coecke \cite{Abramsky}. Of course, we will obtain the same computational results in these two different definitions.

In the category $\mathfrak{F}$, we have the relations similar to \eqref{fs},
\begin{equation}
Q_+\circ\psi_0 \cong \psi_1, \qquad Q_-\circ\psi_1 \cong \psi_0.
\end{equation}
The 2-morphisms of the states in this category are
\begin{eqnarray}
&&\begin{tikzpicture}[>=angle 60,thick,baseline=0pt,scale=0.6]
    \draw (0,0) node {$0$};
    \draw (1,1) arc(180:360:.5);
    \draw (1,1) to (1,2);
    \draw[->] (2,1) to (2,2);
    \draw (3,1) arc(180:360:.5);
    \draw (3,1) to (3,2);
    \draw[->] (4,1) to (4,2);
    \draw (3,-1) arc(180:0:.5);
    \draw[->] (3,-1) to (3,-2);
    \draw (4,-1) to (4,-2);
    \draw[loosely dotted,line width=1pt] (4.5,1.5) to (5.5,1.5);
    \draw[loosely dotted,line width=1pt] (4.5,-1.5) to (5.5,-1.5);
    \draw (6,1) arc(180:360:.5);
    \draw (6,1) to (6,2);
    \draw[->] (7,1) to (7,2);
    \draw (6,-1) arc(180:0:.5);
    \draw[->] (6,-1) to (6,-2);
    \draw (7,-1) to (7,-2);
    \draw (8,0) node {$0$};
    \draw (9.5,0) node {,};
    \draw (12,0) node {$0$};
    \draw[yscale=-1] [postaction={decoration={markings,mark=at position 0 with {\arrowreversed[]{angle 60}}}, decorate}] (13.3,0) circle (.8);
    \draw (14.6,0) node {$0$};
    \useasboundingbox (-.1,-2.1) rectangle (14.7,2.1);
\end{tikzpicture}
\nonumber\\[.5cm]
&&\qquad\qquad\begin{tikzpicture}[>=angle 60,thick,baseline=0pt,scale=0.6]
    \draw (-1,0) node {$1$};
    \draw (1,1) arc(180:360:.5);
    \draw (1,1) to (1,2);
    \draw[->] (2,1) to (2,2);
    \draw[->] (2,-2) .. controls (2,-.5) .. (1,0) .. controls (0,.5) .. (0,2);
    \draw (3,1) arc(180:360:.5);
    \draw (3,1) to (3,2);
    \draw[->] (4,1) to (4,2);
    \draw (3,-1) arc(180:0:.5);
    \draw[->] (3,-1) to (3,-2);
    \draw (4,-1) to (4,-2);
    \draw[loosely dotted,line width=1pt] (4.5,1.5) to (5.5,1.5);
    \draw[loosely dotted,line width=1pt] (4.5,-1.5) to (5.5,-1.5);
    \draw (6,1) arc(180:360:.5);
    \draw (6,1) to (6,2);
    \draw[->] (7,1) to (7,2);
    \draw (6,-1) arc(180:0:.5);
    \draw[->] (6,-1) to (6,-2);
    \draw (7,-1) to (7,-2);
    \draw (8,0) node {$0$};
    \useasboundingbox (-1.1,-2.1) rectangle (8.1,2.1);
  \end{tikzpicture}
\end{eqnarray}
Obviously, we have the orthonormal relation
\begin{equation}
\langle \psi_{n'}, \psi_n\rangle=\mathrm{dim(Hom}_{\mathfrak{F}}(\psi_{n}, \psi_{n'}))= \delta_{n,n'}.~~~~(n,n'=0,1)
\end{equation}
This is exactly the relation \eqref{orth}.

\section{The 2-representation of the fermion algebra}\label{sec5}
In this section, we will construct the 2-representation of the 1D fermion algebra.
Let $R_0$ be some field, here we just choose $R_0=\mathbb{C}$, and $R_1:=M_n(R_0)$ for some $n>1$ is the matrix ring over $R_0$.
Consider the category $R_0\mathrm{-mod}$ of left modules over $R_0$, and the category $R_1\mathrm{-mod}$ of left-modules over $R_1$.
We know that $R_0$ and $R_1$ are Morita equivalent, and the module categories $R_0\mathrm{-mod}$ and $R_1\mathrm{-mod}$ are equivalent.

Let $M$ be the set of all $n\times 1$-matrices whose entries are elements of $R_0$ and $N$ be the set of all $1\times n$-matrices, then $M$, $N$ can be considered as the bimodules ${}_{R_1}M_{R_0}$, ${}_{R_0}N_{R_1}$, respectively. The tensor product of the elements of the bimodules is just the multiplication of matrices.
Obviously, we have $N\otimes_{R_1} M\cong R_0$ as $(R_0,R_0)$-bimodules and $M\otimes_{R_0} N\cong R_1$ as $(R_1,R_1)$-bimodules.
So there are maps
\begin{equation}\label{fg1}
f_0: N\otimes_{R_1} M\to R_0,\qquad g_0: R_0\to N\otimes_{R_1} M
\end{equation}
which satisfy $f_0 g_0 =\mathrm{id}_{R_0}$ and $g_0 f_0 = \mathrm{id}_{N\otimes_{R_1} M}$,
and
\begin{equation}\label{fg2}
f_1: M\otimes_{R_0} N\to R_1,\qquad g_1: R_1\to M\otimes_{R_0} N
\end{equation}
which satisfy $f_1 g_1 =\mathrm{id}_{R_1}$ and $g_1 f_1 = \mathrm{id}_{M\otimes_{R_0} N}$.

We define the functor
\begin{equation}
F_+: R_0\mathrm{-mod}\to R_1\mathrm{-mod}
\end{equation}
of tensoring with the bimodule ${}_{R_1}M_{R_0}$ and the functor
\begin{equation}
F_-: R_1\mathrm{-mod}\to R_0\mathrm{-mod}
\end{equation}
of tensoring with the bimodule ${}_{R_0}N_{R_1}$.
The identity endomorphism of the functor $F_+$ is denoted by the diagram
\begin{equation}
\begin{tikzpicture}[>=angle 60,thick,baseline=0pt]
  \draw (-.8,0) node {$1$};
  \draw (0,-.6) -- (0,.6) [->] ;
  \draw (.8,0) node {$0$};
\end{tikzpicture}
\end{equation}
where the regions labelled by ``$0$'', ``$1$'' mean the categories $R_0\mathrm{-mod}$, $R_1\mathrm{-mod}$, respectively,
and the identity endomorphism of the functor $F_-$ is denoted by
\begin{equation}
\begin{tikzpicture}[>=angle 60,thick,baseline=0pt]
  \draw (-.8,0) node {$0$};
  \draw (0,-.6) -- (0,.6) [<-] ;
  \draw (.8,0) node {$1$};
\end{tikzpicture}
\end{equation}

Obviously, the functors $F_+$, $F_-$ correspond to the 1-morphisms $Q_+$, $Q_-$ in the 2-category $\mathfrak{F}$ respectively.
The relations $N\otimes_{R_1} M\cong R_0$ and $M\otimes_{R_0} N\cong R_1$ are just the isomorphic relations $Q_{-+}\cong \mathrm{id}_{0}$\,, $Q_{+-}\cong \mathrm{id}_{1}$ which we obtained from the string diagrams in previous sections.

We may define the following bimodule maps whose diagrams correspond to the four $U$-turns:
\begin{eqnarray}
\begin{tikzpicture}[>=angle 60,thick,baseline=0pt]
  \draw (.8,-.3) node {$1$};
  \draw [<-](0,-.4) arc(180:0:.8);
  \draw (2,.3) node {$0$};
  \useasboundingbox (-.1,0);
\end{tikzpicture} & &
\begin{array}{lcl} N\otimes_{R_1} M & \xrightarrow[]{~~f_0~~} &  R_0, \\
 h\otimes g & \longmapsto & h g , \quad (g\in{}_{R_1}M_{R_0},~ h\in{}_{R_0}N_{R_1}) \end{array} \label{mp1}\\
\begin{tikzpicture}[>=angle 60,thick,baseline=0pt]
  \draw (2,-.3) node {$1$};
  \draw [<-](0,.4) arc(180:360:.8);
  \draw (.8,.3) node {$0$};
  \useasboundingbox (-.1,0);
\end{tikzpicture} & &
\begin{array}{lcl} R_1 & \xrightarrow[]{~~g_1~~} & M\otimes_{R_0} N, \\
1 & \longmapsto & \sum e_i\otimes e^i, \end{array}\label{mp2} \\
\begin{tikzpicture}[>=angle 60,thick,baseline=0pt]
  \draw (.8,-.3) node {$0$};
  \draw [->](0,-.4) arc(180:0:.8);
  \draw (2,.3) node {$1$};
\end{tikzpicture} & &
\begin{array}{lcl} M\otimes_{R_0} N & \xrightarrow[]{~~f_1~~} &  R_1, \\
 g\otimes h & \longmapsto & g h, \quad (g\in{}_{R_1}M_{R_0},~ h\in{}_{R_0}N_{R_1})\end{array} \\
\begin{tikzpicture}[>=angle 60,thick,baseline=0pt]
  \draw (2,-.3) node {$0$};
  \draw [->](0,.4) arc(180:360:.8);
  \draw (.8,.3) node {$1$};
\end{tikzpicture} & &
\begin{array}{lcl} R_0 & \xrightarrow[]{~~g_0~~} & N\otimes_{R_1} M, \\
1 & \longmapsto & \frac{1}{n}\sum e^i\otimes e_i  , \end{array}
\end{eqnarray}
where $\{e_i\}_{i=1,\ldots n}$ and $\{e^i\}_{i=1,\ldots n}$ are the generators of $M$ and $N$, respectively, and the multiplication of $g, h\in M, N$ is just the same as that in linear algebra.

The relations $g_0 f_0 = \mathrm{id}_{N\otimes_{R_1} M}$, $f_0 g_0 =\mathrm{id}_{R_0}$ and $g_1 f_1 = \mathrm{id}_{M\otimes_{R_0} N}$, $f_1 g_1 =\mathrm{id}_{R_1}$ just correspond to the following diagrammatic relations
\begin{equation}\label{id1}
\begin{tikzpicture}[>=angle 60,thick,baseline=0pt]
  \draw (0,.75) arc (180:360:.5) ;
  \draw (0,1) -- (0,.75) ;
  \draw (1,1) -- (1,.75) [<-];
  \draw (1,-.75) arc (0:180:.5) ;
  \draw (1,-1) -- (1,-.75) ;
  \draw (0,-1) -- (0,-.75) [<-];
  \draw (1.5,0) node {$0$};
  \draw (2.25,0) node {$=$};
  \draw (3,-1) -- (3,1) [<-];
  \draw (4,-1) -- (4,1) [->];
  \draw (4.5,0) node {$0$};
  \draw (5.5,0) node {,};
  \draw[yscale=-1] [postaction={decoration={markings,mark=at position 0 with {\arrowreversed[]{angle 60}}}, decorate}] (7.5,0) circle (.5);
  \draw (8.5,0) node {$0$};
  \draw (9.25,0) node {$=$};
  \draw (10,0) node {$\mathrm{id}_{R_0}$};
  \useasboundingbox (-.2,-1.1) rectangle (10.2,1.1);
\end{tikzpicture}
\end{equation}

\begin{equation}\label{id2}
\begin{tikzpicture}[>=angle 60,thick,baseline=0pt]
  \draw (0,.75) arc (180:360:.5) ;
  \draw (0,1) -- (0,.75) [<-] ;
  \draw (1,1) -- (1,.75);
  \draw (1,-.75) arc (0:180:.5) ;
  \draw (1,-1) -- (1,-.75) [<-];
  \draw (0,-1) -- (0,-.75);
  \draw (1.5,0) node {$1$};
  \draw (2.25,0) node {$=$};
  \draw (3,-1) -- (3,1) [->];
  \draw (4,-1) -- (4,1) [<-];
  \draw (4.5,0) node {$1$};
  \draw (5.5,0) node {,};
  \draw [postaction={decoration={markings,mark=at position 0 with {\arrowreversed[]{angle 60}}}, decorate}] (7.5,0) circle (.5);
  \draw (8.5,0) node {$1$};
  \draw (9.25,0) node {$=$};
  \draw (10,0) node {$\mathrm{id}_{R_1}$};
  \useasboundingbox (-.2,-1.1) rectangle (10.2,1.1);
\end{tikzpicture}
\end{equation}
These are just the relations \eqref{2lr1} and \eqref{2lr2}.

From the relations \eqref{mp1}, \eqref{mp2}, \eqref{id1} and \eqref{id2}, we may also obtain the following relations,
\begin{equation}\label{zz1}
\begin{tikzpicture}[>=angle 60,thick,baseline=0pt]
  \draw (0,0) .. controls (0,1) and (.75,1) .. (.75,0);
  \draw (0,-1) -- (0,0);
  \draw (1.5,0) .. controls (1.5,-1) and (.75,-1) .. (.75,0);
  \draw (1.5,0) -- (1.5,1) [->] ;
  \draw (2,0) node {$0$};
  \draw (2.75,0) node {$=$};
  \draw (3.5,-1) -- (3.5,1) [->] ;
  \draw (4,0) node {$0$};
  \draw (4.75,0) node {$=$};
  \draw (5.5,0) -- (5.5,1) [->] ;
  \draw (5.5,0) .. controls (5.5,-1) and (6.25,-1) .. (6.25,0);
  \draw (6.25,0) .. controls (6.25,1) and (7,1) .. (7,0);
  \draw (7,-1) -- (7,0);
  \draw (7.5,0) node {$0$};
  \useasboundingbox (-.2,-1.1) rectangle (7.7,1.1);
  \end{tikzpicture}
\end{equation}
\begin{equation}
\begin{tikzpicture}[>=angle 60,thick,baseline=0pt]
  \draw (0,0) .. controls (0,1) and (.75,1) .. (.75,0);
  \draw (0,-1) -- (0,0) [<-] ;
  \draw (1.5,0) .. controls (1.5,-1) and (.75,-1) .. (.75,0);
  \draw (1.5,0) -- (1.5,1);
  \draw (2,0) node {$1$};
  \draw (2.75,0) node {$=$};
  \draw (3.5,-1) -- (3.5,1) [<-] ;
  \draw (4,0) node {$1$};
  \draw (4.75,0) node {$=$};
  \draw (5.5,0) -- (5.5,1);
  \draw (5.5,0) .. controls (5.5,-1) and (6.25,-1) .. (6.25,0);
  \draw (6.25,0) .. controls (6.25,1) and (7,1) .. (7,0);
  \draw (7,-1) -- (7,0) [<-] ;
  \draw (7.5,0) node {$1$};
  \useasboundingbox (-.2,-1.1) rectangle (7.7,1.1);
  \end{tikzpicture}
\end{equation}
It is easy to verify these relations. For example, the second equality in \eqref{zz1} is just the relation
\begin{equation}
M\cong R_1\otimes_{R_1} M \to(M\otimes_{R_0} N)\otimes_{R_1} M\cong M\otimes_{R_0} (N\otimes_{R_1} M)\to M \otimes_{R_0} R_0 \cong M,
\end{equation}
and for the elements, we have
\begin{equation}
g=1\otimes g=\sum_i e_i\otimes e^i\otimes g=\sum_i e_i\otimes e^i\otimes \sum_j g_j e_j =\sum_{i,j} g_j e_i\otimes \delta_{ij} =g.
\end{equation}
The first equality in \eqref{zz1} is just the relation
\begin{equation}
\begin{tikzpicture}[>=angle 60,thick,baseline=0pt,scale=0.6]
  \draw (2,0) -- (2,2.5) [->];
  \draw (2,0) .. controls (2,-1.5) and (1,-1.5) .. (1,0) .. controls (1,1.5) and (0,1.5) .. (0,0);
  \draw (0,0) -- (0,-2.5);
  \draw (3,0) node {$0$};
  \draw (4.5,0) node {$=$};
  \draw (6,1) .. controls (6,0) and (7,0) .. (7,1) .. controls (7,2) and (8,2) .. (8,1);
  \draw (8,-1) -- (8,1);
  \draw (8,-1) .. controls (8,-2) and (7,-2) .. (7,-1) .. controls (7,0) and (6,0) .. (6,-1);
  \draw (6,-1) -- (6,-2.5);
  \draw (6,1) -- (6,2.5) [->] ;
  \draw (9,0) node {$0$};
  \draw (10.5,0) node {$=$};
  \draw (12,-2.5) -- (12,2.5) [->] ;
  \draw[yscale=-1.5] [postaction={decoration={markings,mark=at position 0 with {\arrowreversed[]{angle 60}}}, decorate}] (13.4,0) circle (.6);
  \draw (15,0) node {$0$};
  \draw (16.5,0) node {$=$};
  \draw (18,-2.5) -- (18,2.5) [->] ;
  \draw (19,0) node {$0$};
  \useasboundingbox (-.1,-2.6) rectangle (19.1,2.6);
  \end{tikzpicture}
\end{equation}

We may define a 2-category $\mathcal{C}$ as follows,
\begin{itemize}
  \item $\mathrm{Ob} \mathcal{C} = 0,1$.

  \item The 1-morphisms from $i$ to $j$ are functors from $R_i$-mod to $R_j$-mod that are direct summands of compositions of the functors $F_+$, $F_-$.

  \item The 2-morphisms are natural transformations of functors.
\end{itemize}
We may also define a 2-functor $\mathbf{F} : \mathfrak{F} \to \mathcal{C}$ in the obvious way,
\begin{itemize}
  \item For $i \in \mathrm{Ob} \mathfrak{F}$, $\mathbf{F}(i) = i$.

  \item On 1-morphisms, $\mathbf{F}$ maps $Q_\varepsilon \in \mathrm{Hom}_{\mathfrak{F}}(i,j)$ to the tensor product of the bimodules $M$, $N$, where each $+$ corresponds to the bimodule $M$ and each $-$ corresponds to the bimodule $N$.

  \item On 2-morphisms, $\mathbf{F}$ maps a string diagram to the corresponding bimodule map (or, more precisely, to the natural transformation corresponding to this bimodule map) according to the definitions given in this section.
\end{itemize}
The functor $\mathbf{F}$ is a categorification of the Fock space representation of the fermion algebra.

\section{Conclusions and discussions}\label{sec6}
In this paper, based on the diagrammatic methods developed in \cite{Lauda, Khovanov}, we studied the categorification of the 1D fermion algebra.
We constructed a graphical category corresponding to the fermion algebra, and then investigate the properties of this category. We found that in this graphical category, all the string diagrams with crossings are equal to zero. This is quite different from the case in the diagrammatic categorification of the boson algebra \cite{Khovanov}, and this property makes the planer diagrams much simpler than those in the graphical categories corresponding to the boson algebra. Based on this category, we also defined the corresponding 2-category. The categorical analogue of the Fock states are some kind of 1-morphisms in our 2-category, and the dimension of the vector space of 2-morphisms is just the inner product of the corresponding Fock states.
We found that the results in our categorical framework coincide exactly with those in normal quantum mechanics. These convince us that the 2-category we have constructed is the right categorical correspondence of the fermion algebra.

In fact, the fermion algebra can be considered as a special case of the quon algebra when $q=-1$ \cite{Chaichian},
\begin{eqnarray}
&&[\hat{a},\hat{a}^{\dag}]_q:=\hat{a} \hat{a}^{\dag}- q\hat{a}^{\dag}\hat{a}=1,\nonumber\\
&&[\hat{a},\hat{a}]_q=[\hat{a}^{\dag},\hat{a}^{\dag}]_q=0.
\end{eqnarray}
Note that the second relation above is only available when $q=\pm1$. So one may
study the categorification of the fermion algebras via the categorification of the quon algebras \cite{clw}.
In Ref.~\cite{clw}, the authors constructed a $\mathbb{Z}$-graded 2-category corresponding to the quon algebra, and the grading shift in the 2-category just corresponds to the multiplication by $q$ on the Grothendieck group. So one may obtain the fermion algebra if choose the parameter $q=-1$ in the process of decategorification of the graded 2-category.
These two categorification approaches can both obtain the correct categorical correspondences to the fermion algebra. The categorical framework in the approach discussed in the present paper is much closer to the content in normal quantum mechanics. But in the approach discussed in \cite{clw}, the categorifications of the boson algebra and the fermion algebra can be combined into one categorical framework.

All these results of the categorification of the 1D fermion algebra can be easily extended to the case of the categorifications of the higher-dimensional
fermion algebras.
Since the boson and fermion algebras are the simplest and most fundamental algebraic relations in quantum physics, we hope that our results will help us to study the categorification of other physical theories. For example, one can use the results in the present paper to study the categorification of the supersymmetry algebras. Work on this direction is in progress.

\section*{Acknowledgements}
The authors gratefully acknowledge the support of the Morningside Center of Mathematics and Professor Shi-Kun Wang.
This project is partially supported by the National Natural Science Foundation of China under Grant Nos. 10975102, 10871135, 11031005, 11075014.


\begin{thebibliography}{100}

\bibitem{Baez} J. Baez and J. Dolan, ``Categorification.'' \textit{Contemp. Math.} {\bf 230}, 1-36 (1998), arXiv:math/9802029.

\bibitem{Crane} L. Crane and I. B. Frenkel, ``Four-dimensional topological quantum field theory, Hopf categories, and the
canonical bases.'' \textit{J. Math. Phys.} {\bf 35}, 5136-5154 (1994), arXiv:hep-th/9405183.

\bibitem{Baez1} J. C. Baez and J. Dolan, ``From finite sets to Feynman diagrams.'' in \textit{Mathematics Unlimited - 2001 and Beyond}, Vol. 1,
eds. B. Engquist and W. Schmid, Springer, Berlin (2001), arXiv:math/0004133.

\bibitem{Morton} J. Morton, ``Categorified algebra and quantum mechanics.'' \textit{Theory Appl. Categ.} {\bf 16}, 785-854 (2006), arXiv:math/0601458.

\bibitem{Vicary} J. Vicary, ``A categorical framework for the quantum harmonic oscillator.'' \textit{Int. J. Theor. Phys.} {\bf 47}, 3408-3447 (2008), arXiv:0706.0711.

\bibitem{Heunen} C. Heunen, N. P. Landsman and B. Spitters, ``A topos for algebraic quantum theory.'' \textit{Comm. Math. Phys.} {\bf 291}, 63-110 (2009), arXiv:0709.4364.

\bibitem{Abramsky} S. Abramsky and B. Coecke, ``Categorical quantum mechanics.'' in: \textit{Handbook of Quantum Logic and Quantum Structures: Quantum Logic}, eds. K. Engesser, D.M. Gabbay, and D. Lehmann, Elsevier, Amsterdam, pp. 261-323 (2009), arXiv:0808.1023.

\bibitem{Stirling} S. D. Stirling and Y. S. Wu, ``Braided categorical quantum mechanics I.'' (2009), arXiv:0909.0988.

\bibitem{Isham} C. J. Isham, ``Topos methods in the foundations of physics.'' in \textit{Deep Beauty},
ed. H. Halvorson, Cambridge University Press (2011), arXiv:1004.3564.

\bibitem{K1} M. Khovanov, ``NilCoxeter algebras categorify the Weyl algebra.'' \textit{Comm. Algebra} {\bf 29}, 5033-5052 (2001), arXiv:math/9906166.

\bibitem{Lauda} A. D. Lauda, ``A categorification of quantum sl(2).'' \textit{Adv. Math.} {\bf 225}, 3327-3424 (2010), arXiv:0803.3652.

\bibitem{Khovanov} M. Khovanov, ``Heisenberg algebra and a graphical calculus.'' (2010), arXiv:1009.3295.

\bibitem{Cautis} S. Cautis and A. Licata, ``Heisenberg categorification and Hilbert schemes.'' \textit{Duke Math. J.} {\bf 161}, 2469-2547 (2012), arXiv:1009.5147.

\bibitem{Licata} A. Licata and A. Savage, ``Hecke algebras, finite general linear groups, and Heisenberg categorification.'' \textit{Quantum Topology} {\bf 4}, 125-185 (2013), arXiv:1101.0420.

\bibitem{Wang} N. Wang, \emph{et al.}, ``Categorification of fermions.'' (in prepation).

\bibitem{Lauda1} A. D. Lauda, ``An introduction to diagrammatic algebra and categorified quantum sl(2).'' \textit{Bull. Inst. Math. Acad. Sin.} {\bf 7}, 165-270 (2012), arXiv:1106.2128.

\bibitem{lw} B. S. Lin and K. Wu, ``A categorification of the boson oscillator.'' \textit{Commun. Theor. Phys.} {\bf 57}, 34-40 (2012).

\bibitem{KMS} M. Khovanov, V. Mazorchuk and C. Stroppel, ``A brief review of abelian categorifications.''
\textit{Theory Appl. Categ.} {\bf 22}, 479-508 (2009), arXiv:math/0702746.

\bibitem{Chaichian} M. Chaichian, R. G. Felipe and C. Montonen, ``Statistics of $q$-oscillators, quons and relations to fractional statistics.'' \textit{J. Phys. A: Math. Gen.} {\bf 26}, 4017-4034 (1993), arXiv:hep-th/9304111.

\bibitem{clw} L. Q. Cai, B. S. Lin and K. Wu, ``A diagrammatic categorification of $q$-boson and $q$-fermion algebras.'' \textit{Chin. Phys. B} {\bf 21}, 020201 (2012).


\end{thebibliography}
\end{document}